\documentclass[preprint]{iucr}              
\usepackage{xcolor}
\usepackage{graphicx}
     \journalcode{S}              

\begin{document}                  



\title{An algorithm for the automatic deglitching of x-ray absorption spectroscopy data}


\author[a]{Samuel M.}{Wallace}
\author[b]{Marco A.}{Alsina}
\cauthor[a]{Jean-Fran\c{c}ois}{Gaillard}{jf-gaillard@northwestern.edu}

\aff[a]{Department of Civil and Environmental Engineering, Northwestern University, 2145 Sheridan Road, Evanston, \country{USA}}
\aff[b]{Department of Construction Engineering and Management, University of Talca, Camino Los Niches Km. 1, Curic\'{o}, \country{Chile}}









\maketitle                        

\begin{synopsis}
An algorithm was developed for the automated deglitching of full XAS spectra.
\end{synopsis}

\begin{abstract}
Analysis of x-ray absorption spectroscopy (XAS) data often involves the removal of artifacts or \textit{glitches} from the acquired signal, a process commonly known as \textit{deglitching}. Glitches result either from specific orientations of monochromator crystals or from scattering by crystallites in the sample itself. Since the precise energy — or wavelength — location and the intensity of glitches in a spectrum cannot always be predicted, deglitching is often performed on a per spectrum basis by the analyst. Some routines have been proposed, but they are prone to arbitrary selection of spectral artifacts and are often inadequate for processing large data sets.

Here we present a statistically robust algorithm, implemented as a Python program, for the automatic detection and removal of glitches that can be applied to a large number of spectra. It uses a Savitzky-Golay filter to smooth spectra and the generalized extreme Studentized deviate test to identify outliers. We achieve robust, repeatable, and selective removal of glitches using this algorithm.
\end{abstract}

\section{Introduction}

In x-ray absorption spectroscopy (XAS), glitches correspond to artifacts in a confined energy range, usually a few data points, manifesting as a sharp variation in measured absorption \cite{stern82, bunker2010, calvin2013}. Glitches often arise from multiple-diffraction events within the crystal monochromator used to set the energy of the x-rays, resulting in a significant decrease or rise in the intensity of the beam delivered to the sample \cite{bauchspiess84}. Alternatively, diffraction from other crystalline phases, for instance in diamond anvil cells, may also create glitches, impacting the measurement of the absorption coefficient ($\mu(E)$) \cite{sapelkin02}.

Analysts can minimize the impact of glitches by using best practices at the time of data collection, which includes ensuring uniform sample preparation, confirming the linearity of equipment at the beamline, and detuning the monochromator to reduce the intensity of the multiple diffraction events; however, these practices do not completely eliminate glitches \cite{abe18}. Collecting data free from significant artifacts is particularly challenging in some energy ranges, such as at the Fe K-edge using a Si(111) monochromator \cite{SSRL}. Leaving spurious points in place may interfere with the analysis of XAS data; for instance, glitches may introduce error either in the normalization of the spectrum or the transformation of the extended x-ray absorption fine structure (EXAFS) data into R-space \cite{abe18}, and spectra may be improperly clustered. As such, the processing of spectra may require a deglitching step, where data points corresponding to glitches are removed.

Given the erratic nature of glitches, their removal is commonly based on the judgement of the analyst through visual inspection of the data. For example, the program Athena offers a graphical user interface for glitch removal, where one may either conduct point-and-click identification of spurious data points or set a threshold value based on the post-edge normalization curve outside of which data points are removed \cite{ath_deglitch}. Other programs include the capability to remove data points at specified indices \cite{wellenreuther09} or the option to compare $\mu(E)$ with respect to the incoming incident beam intensity $I0$ \cite{aberdam98}. As a result, these methods require the analyst to manually inspect each XAS spectrum to remove the aberrant values, an approach that is readily applicable to large glitches and small datasets.

However, the growing relevance of time-resolved XAS \cite{bak18} and continuous scanning "quick EXAFS" modes \cite{prestipino11} along with increasing accessibility to laboratory-source XAS instruments \cite{anklamm14, jahrman19} promises large datasets where manual deglitching of spectra becomes impractical. A rapid and robust method for automated deglitching of XAS spectra would directly address this shortcoming. Previous methods with greater potential for automation used the first derivative of the absorption coefficient to identify glitches \cite{zhuchkov01}. While useful, such strategies can result in points adjacent to glitches being erroneously identified, and applicability to the x-ray absorption near edge structure (XANES) region is limited. Moreover, a manual threshold must still be set by the analyst to identify glitches based on visual inspection of the data.

Here, we propose a method for the high-throughput deglitching of XAS data. Outliers are identified in the normalized residuals between original and low-pass-filtered data. A second iteration of this process minimizes the occurrence of type I errors. Through this, we achieve accurate and repeatable removal of glitches from full-spectrum XAS data. We present deglitching examples on XANES and EXAFS spectra, as well as a negative control. Finally, we discuss potential limitations of our method as well as strategies for improvement.

\section{Methods}
The deglitching algorithm was implemented as a computer program written in the Python 3.7 programming language. The packages Scipy \cite{scipy} and Numpy \cite{numpy} were used for all calculations, and Larch \cite{larch} was used for XAS data processing in a Jupyter Notebook environment \cite{jupyter}.

The deglitching program was designed to be compatible with the data structure used by Larch for XAS analysis; the only required input for the deglitching program are data channels corresponding to the energy ($E$) and the absorption coefficient ($\mu(E)$) within a Larch group. Absorption data may or may not be normalized prior to deglitching. Additional parameters may be specified here to tune the deglitching program, though default parameters should work in most circumstances.

All spectra presented here were collected at the bending magnet beamline of the Dow-Northwestern-Dupont Collaborative Access Team (DND-CAT), Sector 5 at the Advanced Photon Source at Argonne National Laboratory. Fe K-edge data is presented for drinking water treatment residual samples collected from drinking water treatment plants in the United States. The other samples include a denture adhesive cream and cobalt (II, III) oxide (Aldrich). Energy was set using a Si(111) double crystal monochromator. Intensity values for the incident beam ($I_0$), the transmitted beam ($I_{T1}$), and the secondary transmitted beam ($I_{T2}$) were collected using Oxford ionization chambers with 29.6 cm path lengths. Fluorescence data were captured using Vortex ME4 silicon drift detectors.

\section{Algorithm}

What follows is a description of the deglitching process, also depicted graphically in Figure 1 and outlined in Figure 2. Data channels for energy and $\mu (E)$ are provided to the algorithm (Figure 1.1). First, the $\mu(E)$ channel is fit with a Savitzky-Golay filter to provide a smoothed representation of the data, $\mu_{SG}(E)$ (Figure 1.2A). The Savitzky-Golay filter uses a least-squares fitting procedure to fit each point over a rolling window of odd length ($w_{SG}$), effectively acting as a lowpass filter \cite{savgol} (Figure 1.2B). Savitzky-Golay window length and the polynomial order of the rolling fit are adjustable parameters set at 9 and 3 respectively by default. The results of the filter are then subtracted from the normalized absorption to compute the residuals between the original and filtered data, $\delta \mu(E)_A$ (Figure 1.3).

\begin{equation}
    \delta \mu(E)_A = \mu(E) - \mu_{SG}(E)_A
\end{equation}

Given that the Savitzky-Golay filter acts as a low-pass filter, $\delta \mu(E)_A$ will reflect the contribution of high frequencies to $\mu(E)$. One expects a normal distribution of residuals should Gaussian noise be the only source of misfit between the data and a fit aiming to represent the true values of the data \cite{nist_resid}. Glitches, being aberrant high-frequency spectral features, will result in residuals that are outliers compared to the rest of the residuals. In simple cases, only glitches will be outliers in these residuals. However, glitches may impact $\delta \mu(E)_A$ at any point within $(w_{SG}/2)-1$ points of a glitch, increasing the variance of these residuals relative to those outside the Savitzky-Golay window. Moreover, the residuals in the absence of the glitch will also vary slightly based on regional features of the XAS spectrum. One can generally expect larger values for $|\delta \mu(E)_A|$ near the absorption threshold, owing both to the wider range of absorbances contained in fitting windows in this region and, in cases where the low-pass filter is overly restrictive, the attenuation of some high-frequency features. Identifying outliers on untreated residuals may lead to the false positive identification of glitches in the XANES region or failure to identify subtle glitches in the EXAFS region as a result.

Residuals must be normalized to identify glitches across the full spectrum. Average-based values, like standard deviation, are strongly impacted by outliers, while median values are comparatively more resistant to outliers. For this reason, a rolling median is found for $|\delta \mu (E)_A|$ to acquire a rolling median absolute deviation. Much like the Savitzky-Golay filter, these median values are calculated from a rolling window, $w_m$. The window size for the median calculation is selected such that, should a glitch be contained within $w_m$, more than half the points in $w_m$ will not have included the glitch in their Savitzky-Golay filter fitting window $w_{SG}$. This may be defined as:
\begin{equation}
    w_m = 2*(w_{SG} + L_g -1) + 1
\end{equation}
Where $w_m$ is the window length for calculating the median absolute deviation, $w_{SG}$ is the Savitzky-Golay filter window length, and $L_g$ is the maximum number of points corresponding to a single glitch. Normalized residuals $r$ are computed by dividing $\delta \mu(E)_A$ by the local median absolute deviation of residuals to account for regional variability in the fit (Figure 1.4):
\begin{equation}
    r_i = \frac{\delta \mu(E)_{A, \ i}}{median|\delta \mu(E)_{A, \ w}|},
    \qquad w \in \left\{ i - \frac{w_{m}-1}{2}, \cdots, i + \frac{w_{m}-1}{2} \right\} 
\end{equation}
Where $r_i$ corresponds to the normalized residual value at $E_i$. At the beginning and end of the data, calculations are performed using truncated windows. Provided Savitzky-Golay parameters that do not filter true signal, these normalized residuals will be normally distributed with a standard deviation of approximately 1.48, corresponding to the conversion from median absolute deviation to standard deviation based on the cumulative distribution function of the standard normal distribution \cite{nist_genesd} (Figure 1.4/1.5).

From the normalized residuals, outliers are mathematically identified using a generalized extreme Studentized deviate test (generalized ESD) (Figure 1.5). The generalized ESD identifies outliers in normally-distributed data when provided a maximum number of outliers and a significance value for the outlier identification \cite{genESD}; these values are set at 10\% of the data points contained in the spectrum and 0.025 by default respectively. The first step of the generalized ESD is to calculate the mean of the dataset. Next, the data point furthest from the mean value is identified. The \textit{test statistic} is calculated by finding the distance of this point from the mean value in terms of number of standard deviations. Using a $t$ distribution at the provided significance level, the maximum acceptable distance from the mean is calculated for a dataset of a given length. This provides a \textit{critical value} which is, again, in units of number of standard deviations from the mean. The most distant point is removed and the calculation is repeated until the maximum number of outliers are reached. If the \textit{critical value} of a given point is greater than the test statistic, that point and all previously analyzed points are identified as outliers.

Outliers identified in this first stage are taken as \textit{candidate points} for glitches. A sufficiently large glitch may result in a poor fit from the Savitzky-Golay filter on points within the same filter window as the glitch (Figure 1.6), so outliers in the first pass (\textit{candidate points}) may not all be glitches. Therefore, \textit{candidate points} are removed from a copy of the data, and $\mu(E)$ at these points is interpolated using a cubic spline (Figure 1.6). The resulting $\mu(E)_{B}$ only differs from $\mu(E)$ at the \textit{candidate points}. A second set of Savitzky-Golay filtered data, $\mu_{SG}(E)_B$, is generated based on this copy of the data using the same parameters as before (Figure 1.7A). From here, the process is repeated, finding the residuals ($\delta \mu(E)_B$) between the original data ($\mu(E)$) and the second Savitzky-Golay filter ($\mu_{SG}(E)_B$), defined as:
\begin{equation}
    \delta \mu(E)_B = \mu(E) - \mu_{SG}(E)_B
\end{equation}
As before, these residuals are normalized by the regional median absolute deviation, this time calculated on $\delta \mu(E)_B$. The window for this second normalization shrinks to $(2*L_g)+1$ points, effectively limiting the maximum number of points in a glitch to $L_g$. This adjustment is made possible by the interpolation of \textit{candidate points}, which minimizes the impact of glitches on the filter's fit for other points within $w_{SG}$. Outliers are identified in $\delta \mu(E)_B$ using the generalized ESD. The outliers identified in this second pass that are within one half the Savitzky-Golay window length, $w_{SG}$, of a candidate point are taken to be glitches and removed.

\section{Results and Discussion}
The deglitching algorithm was tested on a range of elements and data collection modes, as shown in Figure 3. For all data, the deglitching procedure was applied across the full spectrum and used a Savitzky-Golay window length of 9, a significance level of 0.025, and a maximum glitch length of 4, the default values for the program. In each of these circumstances, the algorithm successfully identified and removed glitches across the full spectrum of data without removing non-glitch points. Given the diverse datasets included in these examples - which include XANES glitches, minor EXAFS glitches, and data for various elements collected in both absorbance and fluorescence mode, the default parameters should be adequate for most applications.

Figure 3.1 presents a negative control for analysis: a cobalt (II-III) oxide sample with no apparent glitches. The deglitching algorithm identified no points as glitches. Figure 3.2(A-B) shows data for a sample where the algorithm removed a subtle EXAFS glitch at a high K value. Finally, Figure 3.3(A-C) shows an example of the algorithm simultaneously removing glitches in the XANES and EXAFS regions.

This algorithm provides repeatable and robust removal of glitches. As such, for glitches that are subtle or in relatively noisy data, parameters may need to be adjusted to identify and remove these points. Several strategies may be used in these situations. For EXAFS glitches, the low-frequency, low-intensity features of the region may allow stricter low-pass filters to be implemented. A longer Savitzky-Golay filter window will have the effect of filtering more high frequencies from the data. For glitches anywhere in the spectrum, the significance level may be increased for more aggressive outlier identification. For both of these changes, it is recommended to limit the deglitching algorithm to a region of interest that includes the glitch; otherwise, non-glitch points may be removed.

\ack{We thank Dr. Qing Ma for his technical assistance while performing XAS experiments at the APS. Portions of this work were performed at the DuPont-Northwestern-Dow Collaborative Access Team (DND-CAT) located at Sector 5 of the Advanced Photon Source (APS). DND-CAT is supported by Northwestern University, The Dow Chemical Company, and DuPont de Nemours, Inc. This research used resources of the Advanced Photon Source, a U.S. Department of Energy (DOE) Office of Science User Facility operated for the DOE Office of Science by Argonne National Laboratory under Contract No. DE-AC02-06CH11357.}

\bibliographystyle{iucr}
\referencelist[deglitching_manuscript]

\begin{figure}
\caption{The deglitching procedure, outlined textually (left) and graphically (right). First (Panel A), high frequencies in the data (1) are attenuated with a Savitzky-Golay filter (2A; process visualized in Panel B/2B), and the residuals between these are computed (3). These residuals are normalized (4, Panels A and C), and outliers are identified (5, Panel C). Next (Panel D), the outlying points are interpolated (6), and the Savitzky-Golay filter is repeated with this new data (7A). Finally, outliers from the normalized residuals (7B) of the new fit and the original data are identified as glitches and removed. Panel E shows the resulting EXAFS data after deglitching (8).}
\includegraphics[width=105mm]{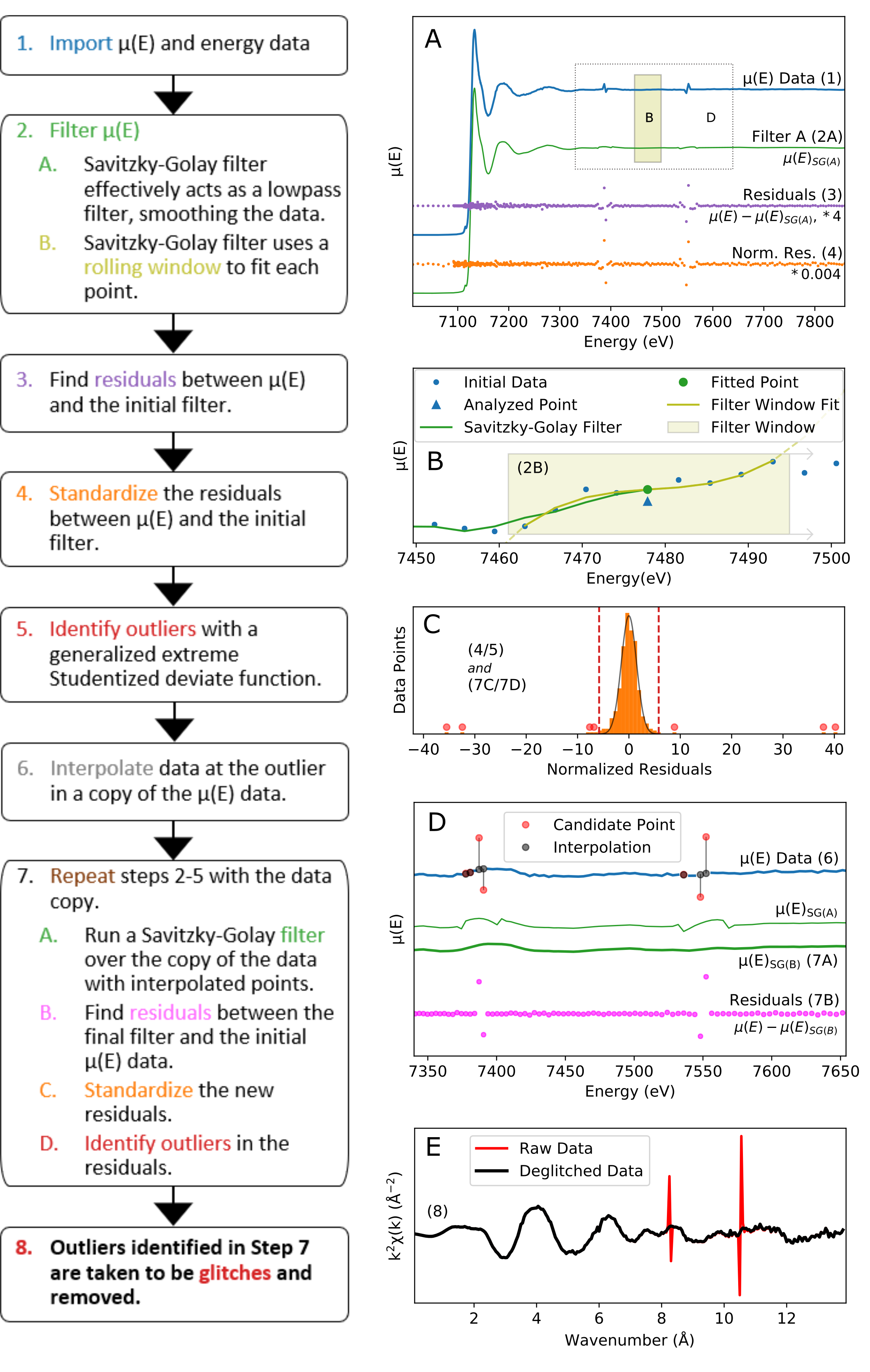}
\end{figure}

\begin{figure}
\caption{A series of XAS spectra are deglitched with our algorithm using default parameters. The left column shows the full spectrum of normalized absorption data, the location of any removed glitches, and the incoming beam intensity I0, with the y-axis scaled to show the range from 77\% to 132\% of the median intensity of I0 for the scan. The right column shows the $k^2$-weighted EXAFS for each spectrum. Row A shows Fe K-edge data collected in fluorescence mode on a drinking water treatment residual sample. Significant monochromator glitches are present in both XANES and EXAFS regions (A1). One point corresponding to a monochromator glitch in the XANES region is removed by the algorithm (A1 inset), and four points corresponding to two separate glitches are removed in the EXAFS region (A2). Row B shows Zn K-edge transmission data collected on a denture adhesive cream (B1). Only one point, corresponding to a monochromator glitch, is removed at approximately 13 \AA$^{1}$ in the EXAFS region (B2). Finally, Row C shows transmission data collected on a cobalt (II-III) oxide sample. Though there is some variation in I0 (C1), there are no obvious effects on either the XANES or the EXAFS (C2), and the algorithm removes no data points.}
\includegraphics[width=90mm]{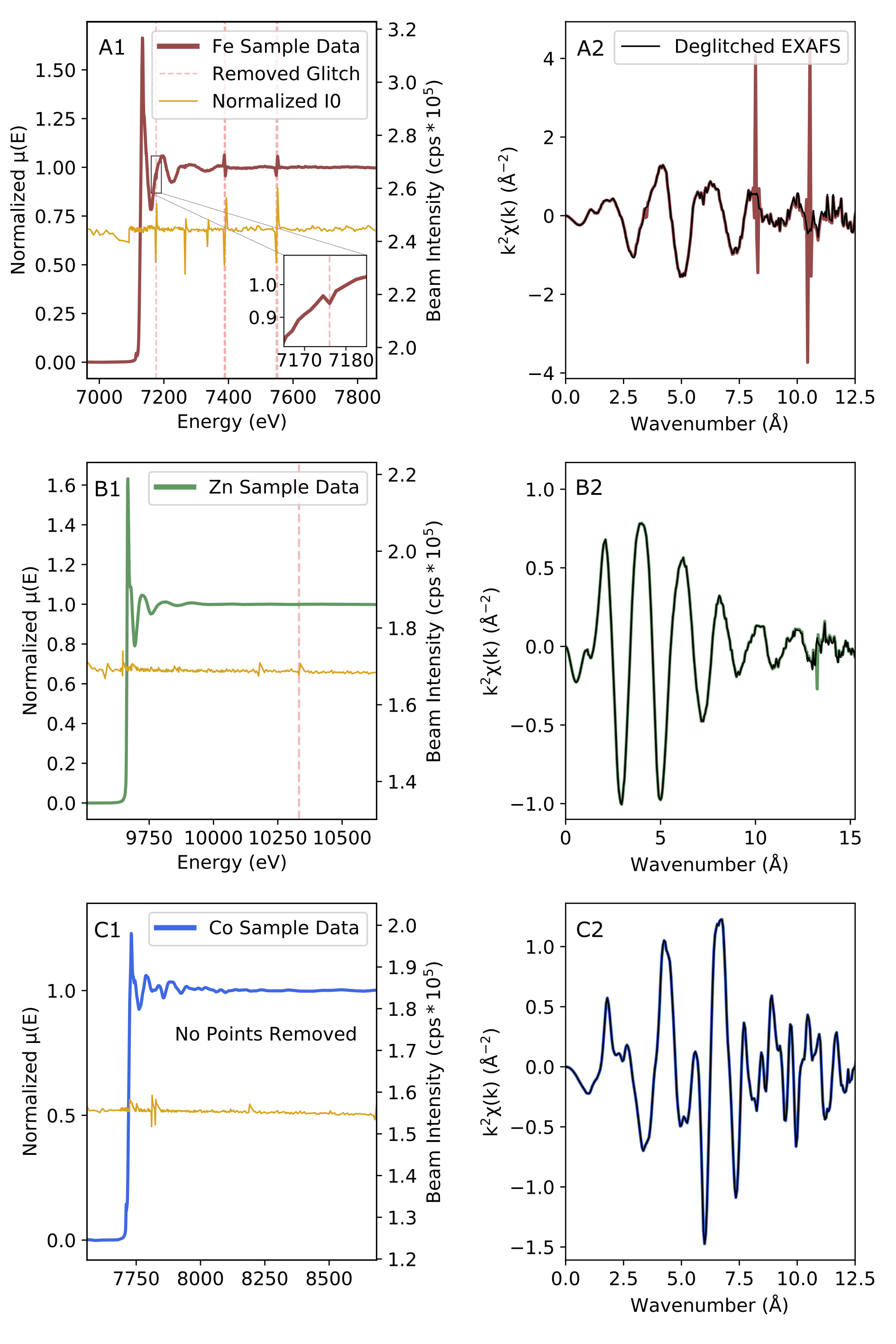}
\end{figure}

\end{document}